\documentclass[11pt,a4paper]{article}
\usepackage{epsfig,float,amsfonts,amssymb}
\begin{document}
\renewcommand{\textfraction}{0.1}
\renewcommand{\thefootnote}{\fnsymbol{footnote}}
\sloppy
\title{\bf Nucleon-Nucleon and Nucleon-Nucleus Optical Models\break  
for Energies to 3 GeV\break
and the Question of NN Hadronization }
\author{H.V. von Geramb$^a$\footnote{geramb@uni-hamburg.de}\ , 
A. Funk$^a$\footnote{funk@physnet2.uni-hamburg.de}, and 
H.F. Arellano$^b$\footnote{arellano@nuclear.dfi.uchile.cl}
\\[0.3cm]
$^a$ Theoretische Kernphysik, Universit\"at Hamburg\\
Luruper Chaussee 149, D-22761 Hamburg, Germany
\\[0.1cm]
$^b$ Departamento de F\'isica, 
Facultad de Cienscias F\'isicas y Matem\'aticas\\
 Universidad de Chile, Cassila 487-3, Santiago, Chile}
\sloppy  
\maketitle         
\begin{abstract} 
Within the key issues of hadronic physics one of the  
interesting issues in nuclear physics is whether there is
a transition region between meson-nucleon and quark-gluon degrees
of freedom in the NN interaction. This question is relevant for pairs of
free nucleons as well as for nucleon pairs immersed in nuclear matter. 
From NN phase shifts we deduce a dibaryonic scale of 1 GeV 
for the soft core  NN potential strengths at nucleon separation 
$0.25<r<0.5$ fm. A short range intermediate transition,  with 
fusion and fission  of the two scattered nucleons into a dibaryon  
with prevailing quark-gluon dynamics,     
is conjectured from NN optical models for $T_{Lab}> 1.5$\,GeV. 
From efforts and progress  of nucleon-nucleus scattering 
analysis in the GeV region some results are presented.  
This is our first step for an in-medium search for transitions 
from the meson-nucleon into the quark-gluon sector using NA 
optical models.

\end{abstract}

\section{Introduction}

A theoretical description of nucleon-nucleon (NN)  scattering  is
a fundamental ingredient  for the understanding of nuclear structure
and scattering  of few- and many-body nuclear systems \cite{Fes92,Mac01}. 
This was and still is a paradigm of nuclear physics. However, 
this scope widens with 
the  intellectual challenge to go beyond the physics 
of a single hadron and understand essential aspects of nuclear physics 
from first principles as the QCD \cite{Rob00}. A recent review of  key 
issues of hadronic physics
comprises the many branches of current fundamental nuclear 
research \cite{Neg00}. 

As the QCD Lagrangian may be considered known, 
it is generally agreed that the nonlinear dynamics of QCD make it
difficult to understand nuclear physics fully from first principles. 
Thus, being less ambitious, it is 
believed that the majority of nuclear physics  phenomena can be 
understood in terms of appropriate
effective degrees of freedom. It is the deductive branch of
nuclear epistemology which builds models, 
suggests methods to determine these degrees of 
freedom  and finds interpretations for them. Being an enthusiast, 
this approach may lead to
emerging new phenomena and  thus to a widened nuclear 
physics realm. All phenomena are associated  with scales and 
the nuclear-hadronic scale 
dependence can be broken into
three regions: nuclear structure at scales $Q\sim 10$\,MeV, the 
hadronic and  dibaryonic 
substructure  scales $Q\sim 1$\,GeV and the partonic region 
scales $Q\gg 1$\,GeV. 
The nuclear energy scale, $Q\sim 10$\,MeV, is very small 
when the natural energy scale of QCD is of the order of
$Q\sim 1$\,GeV. Does this difference  arise from 
cancellations of strongly attractive and repulsive terms 
in the nuclear
interaction or is there some deeper reason for this scale 
change responsible?
Quark-gluon confinement opens a route to a convincing  answer 
of this question.  

The large separation between the hadronic energy scale 
and the nuclear binding scale renders it difficult to apply nonlinear 
QCD directly
to understand the physics of  nuclei. However, quantitative 
calculations based on effective quantum  field theory (EQF) 
techniques that arise
from chiral symmetry provide an alternative approach.
This method has been applied to  pion physics in the 
context of chiral perturbation theory. Currently, it is being 
extended to address few- and many-nucleon interactions. When combined 
with first principle calculations of the low energy constants 
from QCD, these EQF may have the potential to
provide a consistent qualitative understanding of 
low energy properties of nuclei and emergent nuclear structure.
 
In addition to understanding the structure of nuclei from a QCD 
point of view, it is also of interest to understand the 
behavior of nucleons within nuclei.
Well known are modified NN interactions in the form of {\em off-shell 
t- or g-matrices}. They  render useful for nuclear matter 
densities in stable nuclei but this description is
incomplete as deep inelastic scattering experiments by the  EMC
collaboration suggested  that also the quark distribution
in a nucleon,  immersed in  nuclear matter,  differs from
that in free space.
Within the key issues of hadronic physics one of the  interesting 
issues in 
nuclear physics is whether there is
a transition region between meson-nucleon and quark-gluon degrees
of freedom in the NN interaction. This question is relevant for pairs of
free nucleons as well as for nucleon pairs immersed in nuclear matter.
 
In Sect.\,2 we report about our  efforts using inverse scattering 
theory to 
scan  recent NN phase shift analysis to 3 GeV \cite{GWU00} with 
the purpose to 
find precursor phenomena and transitions from  
the meson-nucleon sector into the quark-gluon 
sector (QCD dominated sector). 
To differentiate between reaction mechanism in medium 
energy  NN scattering 
we devised and generated NN optical models for  $0.3<T_{Lab}<3$ 
GeV and calculated
quantities which could reveal intrinsic hadronic excitations 
$\Delta,\,N^\star$ 
as well as fusion and fission of dibaryonic intermediate states. 
A dibaryonic scale of
1 GeV is deduced as NN soft core potential strengths, for NN 
relative distances of
$0.25<r<0.5$ fm. This topic is very  interesting in connection with  
questions of hadronization of the quark-gluon plasma. 

In Sect.\,3  we discuss results of medium energy nucleon-nucleus 
(NA) elastic scattering.
These calculations  use the  NN optical model potentials to generate
 NN t- and g-matrices 
in folding NA optical models. The calculations are based upon 
a relativistically 
corrected full-folding optical model in momentum space that is an 
extension of a 
nonrelativistic predecessor \cite{Are95}. The motivation for these
calculations comes from two opposite directions. The first attempts
a fundamental
understanding of the structure of nuclei from a QCD 
point of view
and the interest to understand the behavior of nucleons within nuclei.
The other direction is given from  applied nuclear
technology where large amounts of nuclear reaction data, including fission 
cross sections at intermediate energies are required  
for accelerator 
transmutation of waste (ATW), for elimination of long-lived 
radioactive wastes with
a spallation source, accelerator-based conversion (ABC) aimed 
to complete the
destruction of weapon plutonium, accelerator-driven energy 
production (ADEP) which
proposes to derive fission energy from thorium with 
concurrent destruction 
of the long-lived waste and without the production of 
nuclear weapon
material, and for accelerator production of tritium (APT) 
\cite{Kal96,Dub97,Mas98,Rip98}.
The results are preliminary and  comprise angular distributions, total and
reaction cross sections for elastic nucleon scattering from the targets
$^{16}$O, $^{40}$Ca, $^{90}$Zr and $^{208}$Pb in an energy
domain  $0.5<T_{Lab}<1.5$\,GeV.
The NA optical model is the best quantitative and microscopic 
nuclear model, based on a
profound understanding of the nuclear many body dynamics 
and NN potentials
in terms of structureless nucleons (protons and neutrons) 
for $E<500$\,MeV 
\cite{Ray92,Amo00}.

\section{Medium energy nucleon-nucleon optical model}

Of the spectrum, low energy NN scattering
traditionally is described in terms of few degrees of
freedom of which  spin and isospin symmetries play the predominant role. 
At medium energies, production processes and inelasticities
become important and several elementary systems composed of nucleons 
and mesons contribute to NN scattering. While these nucleons and
mesons are emergent structures from QCD, at present there is no
quantitative description of NN  scattering above the inelastic threshold
either in terms of QCD or of the emergent nucleons and mesons. 

The experimental NN data and its parameterization
in terms of amplitudes and phase shifts, are very smooth with energy 
to 3 GeV \cite{GWU00,Fun01}; a feature
which supports use of the {\em classic} approach using  
a free NN interaction potential also above 300 MeV.
A  high quality fit of on-shell t-matrices by a potential model is very
desirable  also as it facilitates t- and g-matrix  extension into 
the off-shell domain; 
properties which are needed in  few and many body calculations.
In particular, microscopic optical
model potentials for elastic nucleon-nucleus scattering that 
give quantitative results, 
require a careful and exact treatment of the on- and  
off-shell NN t-matrices 
\cite{Ray92,Are95,Are96,Amo00}. 
Furthermore, calculations of such entities have shown that 
it is crucial to have 
on-shell values of the t-matrices in best possible agreement with 
NN  data at all energies.

There are many studies of few and many body problems
in the low energy regime  $T_{Lab} <300$\,MeV and the results 
have consequences for any model extension above threshold. 
We note in this context that significant off-shell
differences in  t-matrices are known to exist between
the theoretically well motivated boson exchange models of NN 
scattering in this regime. It remains difficult to attribute 
with certainty
any particular dynamical or kinematical feature with those 
differences.
Non-locality, explicit energy dependence and features  associated
with relativistic kinematics are some possibilities.
In contrast, there is the quantum inverse scattering approach by 
which any on-shell t-matrix can be continued into the off-shell domain. 
A specific method is the Gel'fand--Levitan--Marchenko 
inversion algorithm for Sturm--Liouville equations. 
This approach to specify t-matrices off-shell is appropriate when the 
physical S-matrix is unitary and the equation of motion is of the
Sturm--Liouville type. Such  is valid without modification  for 
NN  t-matrices in the energy regime to 300 MeV and for the 
unitary part of the
S-matrix above this energy.

In the spirit of general inverse problems, 
we extended the Gel'fand--Levitan--Marchenko method by additional 
potential terms 
which are determined by a demand to reproduce
perfectly the experimental data (here the partial wave phase 
shift analysis). 
By that means 
NN  optical models,  separately for each partial wave, were generated. 
The  algorithm we  have developed allows studies of
complex local and/or separable potentials in combination with 
any  background  
potential \cite{Ger98,Fun01}. We limited the reference potential  
to the well known real r-space 
potentials from Paris \cite{Lac80}, Nijmegen \cite{Sto93}, 
Argonne \cite{Wir95},
and from Gel'fand--Levitan--Marchenko quantum inversion \cite{Kir89,San97}. 
To them we add channel dependent complex separable potentials with  
energy dependent strengths. For given input data result unique
full optical model potentials that are defined within a 
given potential class.

NN scattering formally is given by the Bethe--Salpeter equation
\begin{equation} \label{eqn_II.1}
{\cal M} = {\cal V} + {\cal V}{\cal G}{\cal M}\ ,
\end{equation}
where  $\cal M$ are invariant amplitudes
that are based upon all connected two particle irreducible diagrams.
This equation serves generally as an ansatz for
approximations. Of those, the three dimensional reductions
are of great use which allow the definition of a potential. 
In particular, the Blankenbecler--Sugar reduction gives an equation
\begin{equation}\label{eqn_II.2}
{\cal T} ({\bf q}^\prime,{\bf q}) =  {\cal V} ({\bf q}^\prime,{\bf q}) +
\int { d^3k \over (2 \pi)^3}  {\cal V} ({\bf q}^\prime,{\bf k})
 { M^2 \over E_k} {1 \over {\bf q}^2 - {\bf k}^2 + i \varepsilon}
{\cal T} ({\bf k},{\bf q}).
\end{equation}
Using  the substitutions \begin{equation}\label{eqn_II.3}
T ({\bf q}^\prime,{\bf q}) = \sqrt{M \over E_{q^\prime}} 
{\cal T} ({\bf q}^\prime,{\bf q})
\sqrt{M \over E_{q}},\
\mbox{\ and}\qquad
V ({\bf q}^\prime,{\bf q})
= \sqrt{M \over E_{q^\prime}}
{\cal V} ({\bf q}^\prime,{\bf q})
\sqrt{ M \over E_{q}},
\end{equation}
a simplified form of the t-matrix is obtained. It is the familiar 
Lippmann--Schwinger equation  
\begin{equation}\label{eqn_II.4}
 T ({\bf q}^\prime,{\bf q}) =  V ({\bf q}^\prime,{\bf q}) +
\int { d^3k \over (2 \pi)^3}  V ({\bf q}^\prime,{\bf k})
 {M \over {\bf q}^2 - {\bf k}^2 + i \varepsilon}
 T ({\bf k},{\bf q})\ .
\end{equation}
The equivalence between the Lippmann--Schwinger
integral equation  and the Schr\"odinger equation gives 
\begin{equation} \label{eqn_II.5}
\left(-\Delta +{M\over \hbar^2} {\cal V}-k^2\right) \psi(\bf{r,k}) = 0.
\end{equation}
The potential $\cal V$ stands representative for  the 
{\em full NN optical model}
which is split into  real, explicitly momentum dependent 
{\em reference potentials},  and 
complex local and  nonlocal {\em  optical model potentials}
\begin{eqnarray} \label{eqn_II.678}
{\cal V}:\, &=&\, V_a(r)+V_b(r)p^2+p^2V_b(r)\nonumber 
\\[0.3cm]& &\qquad\qquad\qquad + V_c(r)+iW_c(r)
\\[0.2cm]& &\qquad\qquad\qquad\qquad 
+ \int ds\, \left(V_d(r,s)+iW_d(r,s)\right).\nonumber 
\end{eqnarray}
The subscripts $a,b,c,d$ stand  representative for 
complete sets of quantum numbers, strengths and range  parameters 
to specify  radial form factors in a partial wave 
decomposition, {\em viz.}
$(a,b,c,d)=(L,L',S,J,T;a_1\cdots a_n;\,b_1\cdots 
b_n;\,c_1\cdots c_n;\,d_1\cdots d_n)$. 
They specify the full optical model and  are specified by: 
simple fitting procedures 
\cite{Ger98}, several high quality boson exchange
potentials OBEP \cite{Lac80,Sto93,Wir95}, 
Gel'fand--Levitan--Marchenko inversion \cite{San97,Fun01}
and generalized inversion algorithm for non-local 
optical potentials \cite{Fun01}.

\subsection{Results of  NN optical model potentials}

A particular solution of potentials $V_{a,b,c,d}$ is $V_a\neq 0$ and
$V_{b,c,d}=0$. Such potential is able to reproduce the real phase shifts
and may be generated with inversion techniques.
There are two equivalent inversion algorithms for the Sturm--Liouville
equation, which one identifies as the Marchenko and the 
Gel'fand--Levitan inversion. 
Both yield principally the same solution and  numerically 
they are  complementary. 
The salient features are outlined for the case of uncoupled 
 Marchenko inversion.

In the Marchenko inversion the experimental information enters via the
S-matrix, $S_{\ell}(k)=\exp(2i\delta_{\ell}(k))$, with which  
an input kernel is defined 
in the form of a Fourier-Hankel transform
\begin{equation} \label{eqn_II.9}
F_\ell (r,t) = -\frac{1}{2\pi} \int_{-\infty}^{+\infty} h^+_\ell(rk)
\left[ S_\ell(k)-1 \right] h^+_\ell(tk) dk,
\end{equation} 
where $h^+_\ell(x)$ are  Riccati-Hankel functions. 
This input kernel when used in the Marchenko equation,  
\begin{equation} \label{eqn_II.10}
A_\ell (r,t)+F_\ell (r,t)+\int_{r}^{\infty}A_\ell(r,s)F_\ell(s,t)ds = 0,
\end{equation}
specifies the translation kernel
$A_{\ell}(r,t)$. The potential 
is a boundary condition for that translational kernel,
\begin{equation} \label{eqn_II.11}
V_{\ell}(r)=-2\frac{d}{dr}A_{\ell}(r,r).
\end{equation}

\begin{figure}[H]
\centering
\epsfig{file=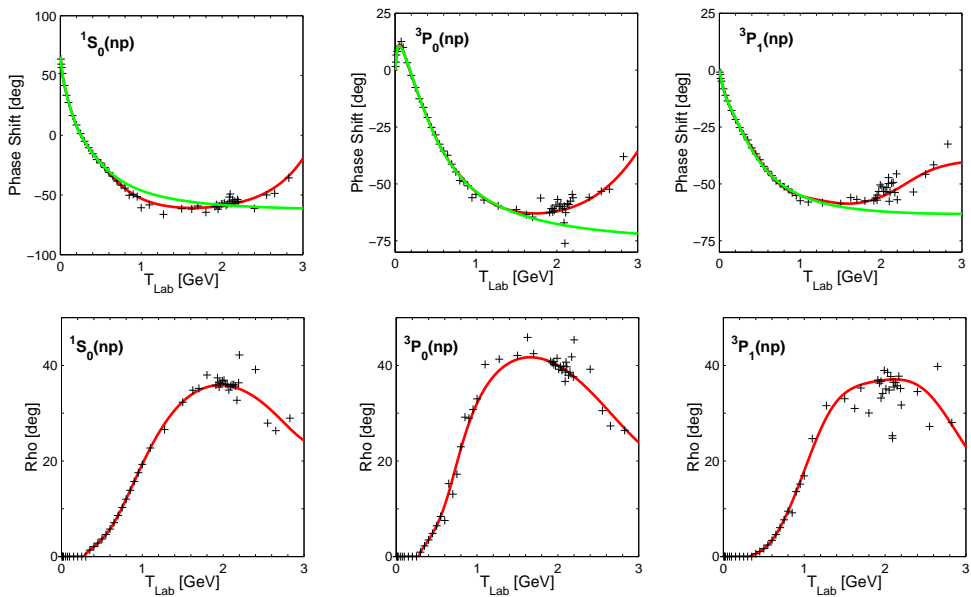,scale=0.5}
\caption{$^1S_0$,$^3P_{0,1}$ SP00 continuous energy 
soltions (solid line,[red]), single energy solutions 
(pluses) and inversion potential phases (grey line,[green])}
\label{Figure_N1}
\end{figure}

\begin{figure}[H]
\centering
\epsfig{file=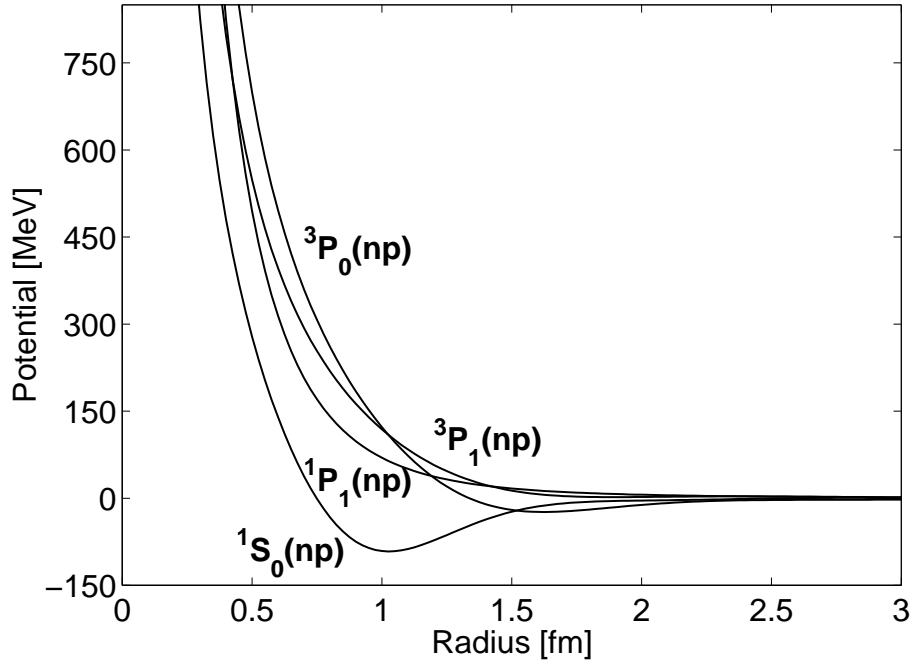,scale=0.75}
\caption{NN inversion potentials based upon  GWU/SP00 phase shifts.}
\label{Figure_N2}
\end{figure}

\begin{figure}[H]
\centering
\epsfig{file=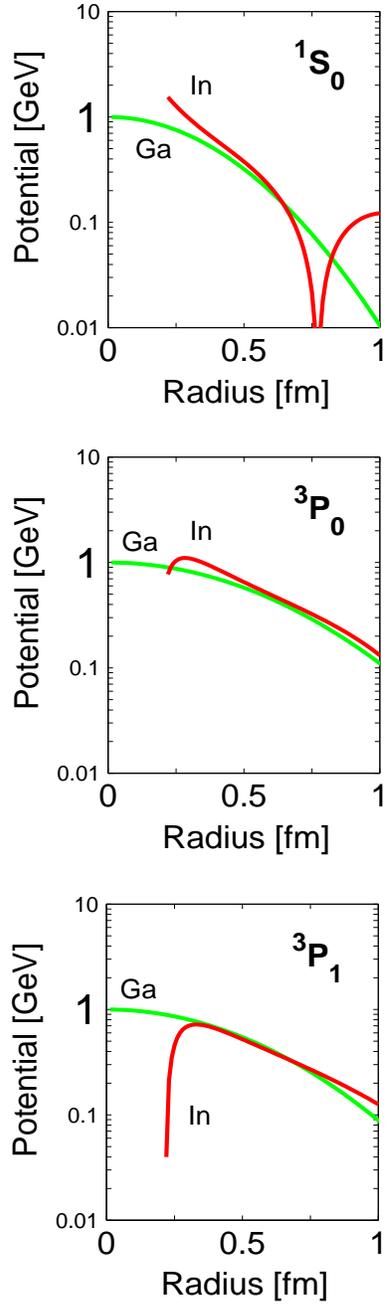,scale=0.65}
\caption{{\em np} $^1S_0$ and $^3P_{0,1}$ inversion potentials 
which reproduce the higher energy GWU/SP00 real phase shifts 
1.2 to 3 GeV particularly well (solid line,[red]). Fitted Gaussian
potentials are also shown  (grey line,[green]).}
\label{Figure_N3}
\end{figure}

Such inversion potentials, $V_a\neq 0$, $V_{b,c,d}=0$, $W_{c,d}=0$,
have been made also to follow closely the GWU/SP00
real phase shifts to 3 GeV \cite{GWU00}. The phase shifts are shown in 
Fig.\,\ref{Figure_N1}, and potentials are displayed in Fig.\,\ref{Figure_N2}.
They possess a long range Yukawa tail, a medium range attraction 
$\sim$1-2 fm and a strong short range repulsion with an onset at 1 fm.
These potentials are energy independent so that the long and
medium range potential properties diminish in 
importance for kinetic  energies
above 500 MeV. For projectiles with $T_{Lab}>1.5$\,GeV essentially
only the repulsive core of these potentials remains of 
significance for scattering. 
Thus inversion potentials have also been obtained with the 
GWU/SP00 real phase shifts optimally fitted 
to 3 GeV \cite{GWU00} using  weighted data, 
$w_{Low}=0.1$ for $T_{Lab}<1.2$\,GeV  
and $w_{High}=1$ for higher energies, to  emphasize the high energy
data and fix more stringently the short range ($<1$ fm) 
character of the deduced interaction. 
The short range properties of these  inversion potentials so found are
displayed in Fig.\,\ref{Figure_N3} and more details are described and
shown in \cite{Fun01}.

\begin{figure}[H]
\centering
\epsfig{file=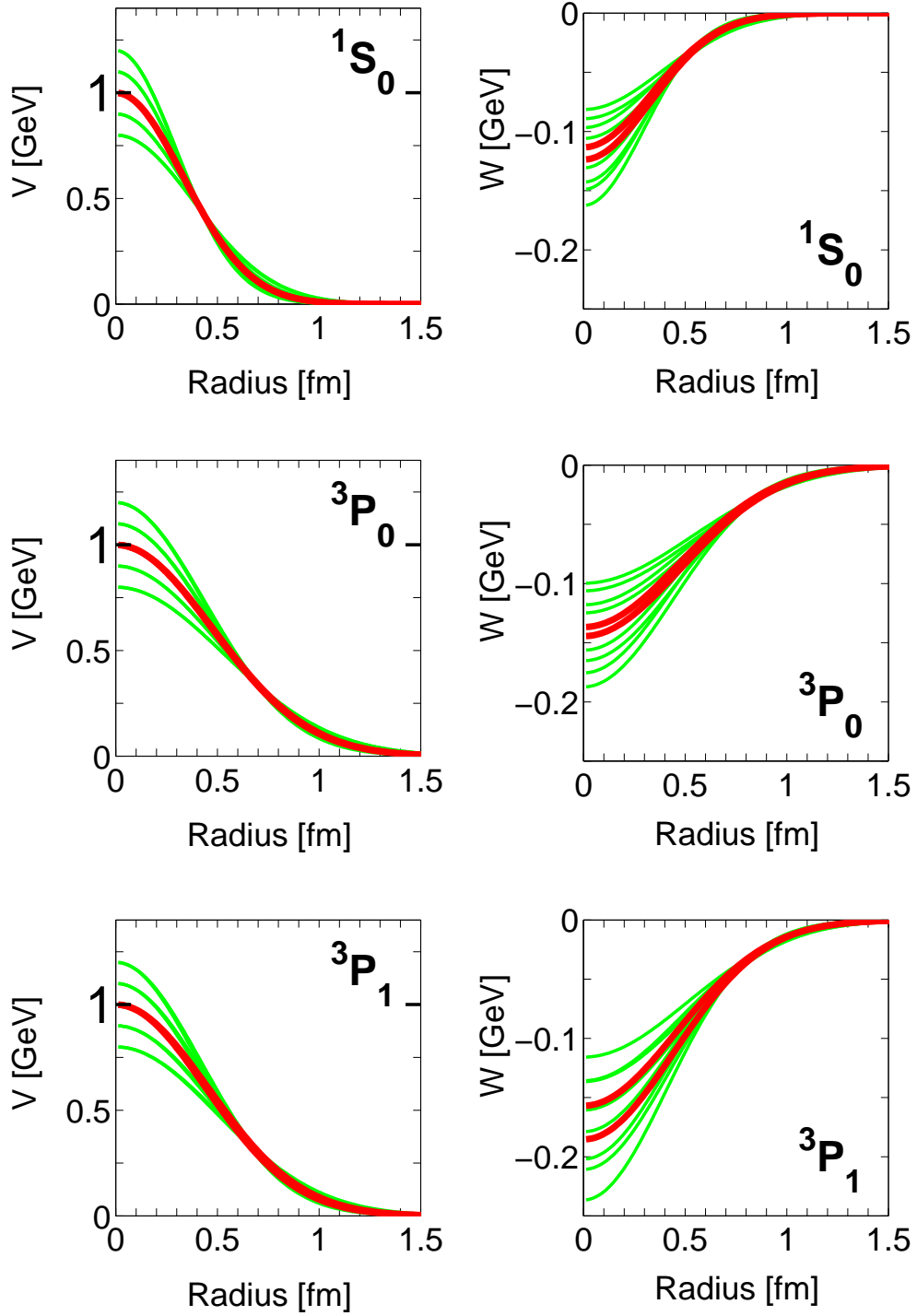,scale=1.0}
\caption{Gaussian optical model  with $V=1$\,GeV fixed
(solid line,[red]),
and free floating for $1.8< T_{Lab} < 2.0$\,GeV (grey lines,[green]).}
\label{Figure_N4}
\end{figure}

 The key features are constrained to low partial
waves $^1S_0$ and $^3P_{0,1}$ (the $^1P_1$ $T=0$ 
{\em np} channel, is limited to  $T_{Lab}<1.3$\,GeV)
for $1.5<T_{Lab}< 2.5$\,GeV. This energy region was also  fitted with a
simple Gaussian potential  to filter and emphasize the
short range region. The higher partial waves are shielded by the
centripetal barrier and they require higher energies to
yield essential fusion/fission contributions. The Gaussian 
potential ansatz restricts  Eqs.\,(\ref{eqn_II.678}) to 
$V_{a,b}=0$, $V_d=0$, $W_d=0$ and
\begin{equation}
V_c+iW_c=\bigl( V(LSJ,E)+iW(LSJ,E) \bigr) e^ {-r^2/r_0^2(LSJ,E)}. 
\label{eqn_II.12}     
\end{equation}
We have fitted the potential strengths and ranges, 
using phase shifts
at many energies within intervals $T_{Lab}-100 <E <T_{Lab}+100$
MeV,  and use $E=T_{Lab}$ in Eq.\,(\ref{eqn_II.678}). It is important 
to notice that the real strength has a robust value $V(LSJ,E)\sim 1$\,GeV
and that the real Gaussian potentials follow closely the
energy independent  Gel'fand--Levitan--Marchenko inversion potentials.
This result is shown in Fig.\,\ref{Figure_N3} and Fig.\,\ref{Figure_N4}.

In Fig.\,\ref{Figure_N4} are shown some of the variations for the
individual channel fits.  The fits  imply  a  robust  
$r_0(^1S_0)=0.46$\,fm,  $r_0(^3P_{0})= 0.66$\,fm and 
$r_0(^3P_{1})= 0.63$\,fm with an imaginary absorptive potential 
which is 25\% stronger in the p- than in the s-channel.

\begin{figure}[H]
\centering
\epsfig{file=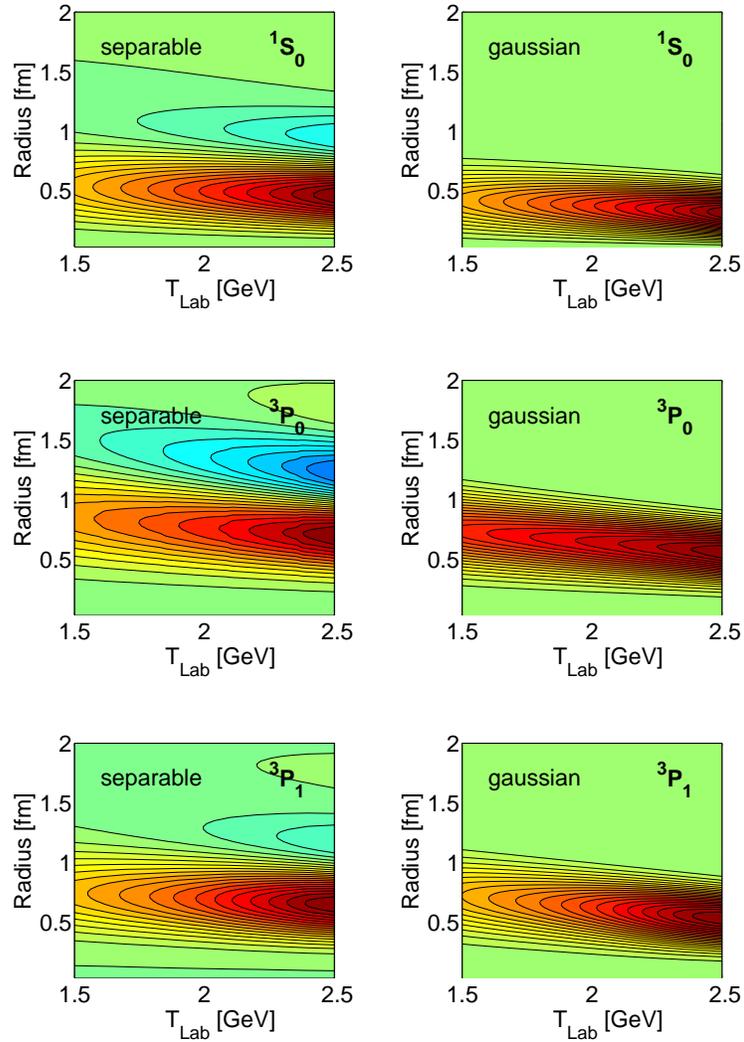,scale=0.55}
\caption{Loss of flux calculated with reference plus separable optical model 
\cite{Fun01}
(left column) and Gaussian optical model (right column) 
for  $1.5<T_{Lab}< 2.5$\,GeV. It is to show that the fusion/fission mechanism 
is practically  the same  for a non-local and  local
potential in  $^1S_0$ and  $^3P_{0,1}$ channels.}
\label{Figure_N5}
\end{figure}

The imaginary potentials account for {\em  flux-loss}
from the elastic scattering channel and this effect is displayed
in the right column of sub-figures in Fig.\,\ref{Figure_N5}. 
The flux-loss was calculated with  normalized
radial wave functions and the imaginary part of the optical model
potentials 
\begin{equation}\label{eqn_II.13}
({\bf \nabla \cdot j})=-{2\over \hbar}{\cal R}\mbox{e}\,\bigl[
\psi^\dagger_{(LSJ,E)}(r,k)W(LSJ,E)e^{-r^2/r_0^2(LSJ,E)}
\psi_{(LSJ,E)}(r,k)\bigr].
\end{equation}
The left sub-column in  Fig.\,\ref{Figure_N5} contains the
results of \cite{Fun01} that use as full optical model, 
a reference (INVS) plus separable optical model potential.
 We confirm with the Gaussian optical model  
the  previously noticed decoupling of the
real and imaginary potentials and the independence of the flux-loss
picture calculated with either a separable (left column) or a local 
(right column) potential. The radial scale of real potential and
absorption is
significantly smaller than the scale of nucleon charge form
factors or a convolution  of two nucleons in a folding approach.

\begin{figure}[H]
\centering
\epsfig{file=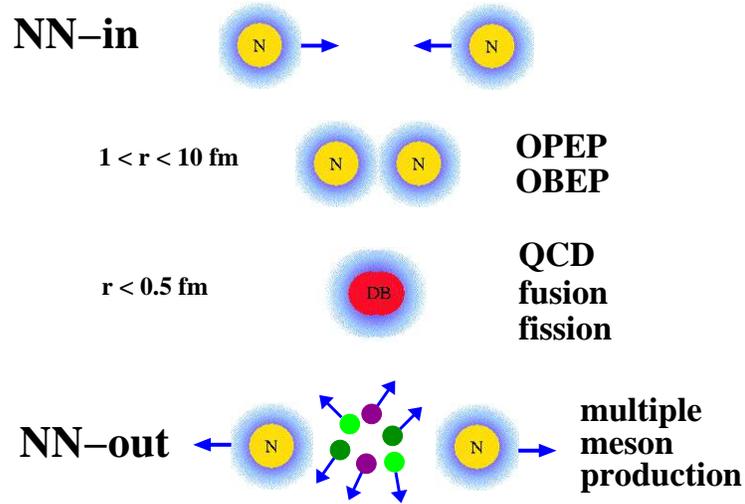,scale=0.8}
\caption{Medium energy NN reaction scheme with intermediate fusion/fission
into a dibaryonic phase and  subsequent decays.}
\label{Figure_N6}
\end{figure}

A reaction scheme for this sequence is shown in Fig.\,\ref{Figure_N6}. 

  Some  further simple optical model studies use  potentials of the form
\begin{equation}
V_c+iW_c=\bigl( V(LSJ,E)+iW(LSJ,E) \bigr) \exp {-r^n/r_0^n(LSJ,E)},\  
\mbox{with}\quad n=1,2,4.
\label{eqn_II.14}     
\end{equation}
All data in the interval $1.5<T_{Lab}<2.5$\,GeV were used and the best static
potential of Eq.\,(\ref{eqn_II.14}) was sought. Table\,1 contains the parameters
of this search. Notice, the $\sim 1$\,GeV real potential strength appears again
but with different form factors. 

\begin{figure}[H]
\centering
\epsfig{file=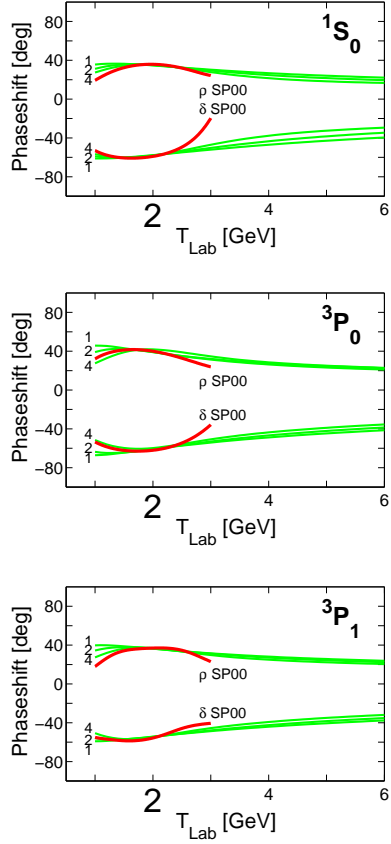,scale=0.5}
\caption{Static Gaussian potential fits around 1.5--2.5 GeV and their
phase shift extrapolations.}
\label{Figure_N7}
\end{figure}

\begin{figure}[H]
\centering
\epsfig{file=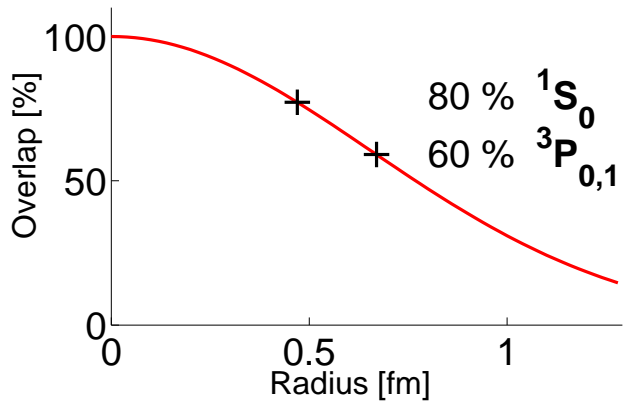,scale=0.6}
\caption{Overlap of nucleon charge form factors.}
\label{Figure_N8}
\end{figure}

\begin{table}\label{Table_III.1}
\centering
\begin{tabular}{c c c c c} \hline\hline\\[-0.2cm]
 Channel & Power   &  $r_0$ [fm]   & \ $V$ [MeV] & \ $W$ [MeV]            \\
\hline\\[-0.2cm]
$^1S_0$  &  1      &  0.26         & 1881   &      -237                   \\
         & {\bf  2}      &  {\bf 0.48}         & {\bf 948}    &      -117 \\
         &  4      &  0.58         & 663    &      -79                    \\
\\[0.1cm]
$^3P_0$  & {\bf  1}      &  {\bf 0.54}         & {\bf 1037}   &      -126 \\
         &  2      &  0.87         & 624    &      -75                    \\
         &  4      &  0.88         & 544    &      -90                    \\
\\[0.1cm]
$^3P_1$  &  {\bf 1}      &  {\bf 0.50}         & {\bf 1029}   &      -171 \\
         &  2      &  0.81         & 609    &      -100      \\
         &  4      &  0.90         & 475    &      -77   
\\[0.1cm]
\hline\hline
\end{tabular}  
\caption{Gaussian ranges and strengths for 
         fits with different powers.}
\end{table}
In Fig.\,\ref{Figure_N7} are shown the
phase shifts $\delta(T_{Lab})$ and $\rho(T_{Lab})$ for the $^1S_0$ 
and $^3P_{0,1}$ channels. 
They  reproduce for any power (n=1,2,4) in Eq.\,(\ref{eqn_II.14})
the experimental minimum in $\delta(T_{Lab})$  and the maximum
in $\rho(T_{Lab})$ at $\sim 1.8-2.0$\,GeV. The trend of data suggests a more
rapid fall-off from the extremum in the phase sifts $\delta$ and $\rho$ but 
more data at higher energies are required and this simple model may guide the
expectations. With the assumption of a Gaussian optical model is implied
a reference to a Gaussian charge form factor of the nucleons.
The overlap of charge form factors of two
nucleons is shown in Fig.\,\ref{Figure_N8}. 

This  calculation helps to visualize  an intermediate  fusion into 
a combined elementary particle (dibaryon) with a similar size as a nucleon. In
particular notice  the overlap of 80\% for the $^1S_0,\, r_0=0.45$\,fm 
and 60\% for the $^3P_{0,1},\,r_0=0.65$\,fm channels.  

A folding approach uses 
\begin{equation}
V_{NN}(|r_1-r_2|))=\int \int ds_1\, ds_2 \,
\rho_N(|r_1+s_1|)\, Q_{qq}(|r_1+s_1-r_2-s_2|)\,\rho_N(|r_2+s_2|)
\end{equation} 
with a quark density in a nucleon  $\rho_N(x)$ and an effective 
interaction $Q_{qq}(s)$ which acts between the continuously 
distributed quark density components. 
This double folding expression may be formulated  with
nonrelativistic/relativistic  quark models and gluon 
exchange as well as QCD based dynamical models with 
effective quark masses and Goldstone boson exchanges. However,  all this
goes beyond what we have done so far but links to
current approaches \cite{Cap00}. For us, it is beyond any doubt and
appears urgent to find a microscopic   
interpretation of the strengths and ranges 
observed with the phenomenological
or inversion optical model potentials.

In summary,  from this analysis we draw the conclusion 
that the GWU phase shift analyses (understood as 
representative for experimental data) \cite{GWU00} 
yield  consistent optical model potentials
with  fluxloss at short distances -- interpreted 
as dibaryonic fusion/fission mechanism --
and a real soft core potential with a strength $\sim 1$\,GeV. This
value requires a deeper fundamental understanding, in particular 
and not the least, 
since the core potential value coincides with the proton and neutron masses.
The formation of an intermediate dibaryon is favoured for the $^1S_0$ and
less favoured for the $^3P_{0,1}$ channels. For higher partial waves
the intermediate individual excitations of the participating nucleons into
$\Delta$,N$^\star$ is prevailing, see medium and high energy reaction schemes, 
Figs. 9 and 10 in \cite{Fun01}.

\section{Nucleon-nucleus  optical model}

\begin{figure}[H]
\centering
\epsfig{file=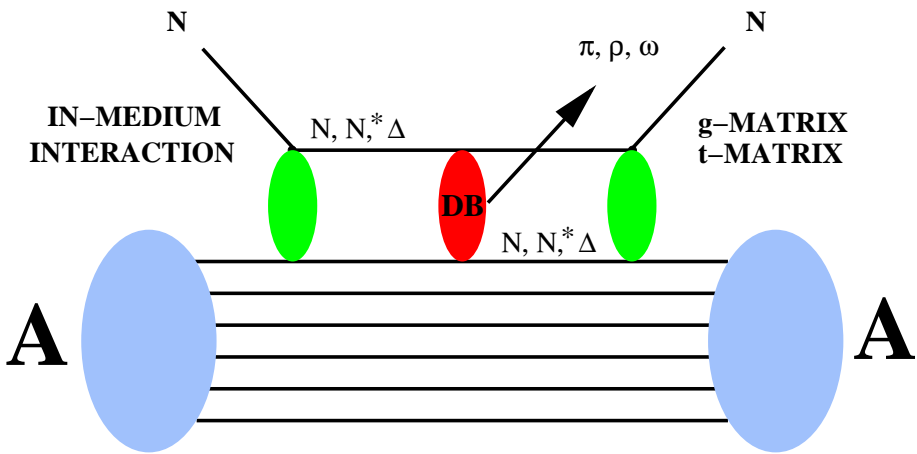,scale=1.25}
\caption{Reaction scheme for NA optical models.}
\label{Figure_A1}
\end{figure}

In Fig.\,\ref{Figure_A1} we show a reaction scheme which underlies all microscopic
optical model approaches. The NA optical model is defined as a folding optical model
in which the {\em in-medium} two particle effective interaction is convoluted with 
the correlated target structure. Principally,  the free NN interaction is used 
to generate any part of  effective {\em in-medium} interaction such as two particle
t- and g-matrices. For energies above the meson production threshold  NN scattering
becomes a three and many body problem which  prohibits a consistent
treatment solely with two nucleons. Instead of incorporating at such high energies
meson channels we continued, {\em in the past}, by neglecting this effect
instead of using a nucleon-nucleon optical model. Now we used an
NN optical model with the effect that the imaginary part 
of the t- and g-matrix has a twofold origin. These are the contributions from
the Greens function pole and the complex NN optical potential. 
We assume that off-shell t- and g-matrices are also consistently described   
with the NN optical model. 

Over the past three decades several efforts have succeeded in
providing a detailed first-order description of intermediate
energy nucleon-nucleus (NA) scattering \cite{Are95,Ray92,Amo00}.
The explicit and accurate treatment of the effective
NN interaction on- and off-shell is undisputed and
in accordance with the full-folding optical model (FF-OMP). In the past, 
emphasis has been placed on retaining a one boson exchange (OBEP)
real  NN interaction  to calculate off-shell t- and g-matrices.
This restricted NA optical model studies to projectile  energies $<350$\,MeV.
Recently, inversion potentials,  calculated  to fit elastic NN 
phase-shift data up to 1.3 GeV in the laboratory frame, have been used in 
FF-OMP calculations for projectile energies $\leq 500$\,MeV \cite{Are96}. 
Although this extension gave  a better description of NA scattering  
relative to previous ones, they still neglected NN absorption and 
an adequate treatment of relativistic kinematics
in the folding procedure and  effects of isobar excitations.
Thus, the primary aim of this work is to extend the range of the  
nonrelativistic FF-OMP into a relativistically corrected version 
which can be used for NA scattering at energies $< 2.5$\,GeV. 

To this purpose one of the interesting issues is whether 
there is evidence of a transition region between meson-nucleon 
and quark-gluon degrees of freedom, which we conjecture in the free NN 
interaction for $T_{Lab}\sim 2$\,GeV,
for nucleon pairs immersed in nuclear matter. 
Answers to this issue are still pending but work 
is in progress and some results are  shown.

Since the Lorentz contraction scales as the ratio of  projectile
energy to mass, and NA scattering at projectile energies $> 300$\,MeV  
have noticeable contraction effects,
it is necessary to include  kinematical and dynamical  
corrections at all stages of the  
FF-OMP. Some relativistic effects are 
{\em effectively} included in the  NN t- and g-matrix
calculations whereas the projectile-nucleus relativistic corrections are 
readily  included in manners which have been developed and 
verified in the past \cite{Ray92}.
Thus, the study of NA scattering with relativistic 
models is not new and several successful approaches are
known \cite{Ray92,Mur87,Ott91,Tjo91,Coo93}.
Consistent models, of the t$\rho$ impulse approximation, treat
kinematical and  dynamical relativistic effects within one framework. 
It was shown for optical models in general, that use of  the
best possible target structure in  combination with 
a first order approximation of the dynamical corrections of 
NA scattering, yield a high quality  microscopic OMP \cite{Amo00}. 
The purpose and result of this work is  
to show that, with relativistic kinematic
corrections in the FF-OMP,  high quality NA scattering results 
for energies $0.5<T_{Lab}<1.5$\,GeV are obtained. 

In the nonrelativistic FF-OMP
the coupling between projectile and mixed ground state density of 
target nucleons
is given by an effective NN interaction and,
in the projectile-nucleus CM frame, the optical potential
is defined by
\begin{equation}
U({\bf k^\prime,  k}, E) = \sum_{\alpha\leq\epsilon_F} 
\int \int d {\bf p}^\prime\, d{\bf p} \, \phi_\alpha^\dagger ({\bf p}^\prime ) 
\langle {\bf k}^\prime {\bf  p}^\prime | {\cal T}(E_\alpha) | 
{\bf k} {\bf  p} \rangle_{A+N} \phi_\alpha({\bf  p})\ .
\label{eqn_III.1}
\end{equation}
$\phi_\alpha $ represents target 
single-particle wave functions of energy $\epsilon_\alpha$,
and $\alpha\leq\epsilon_F$ indicates it limitation below the
Fermi energy $\epsilon_F$. The NN $\cal T$ is calculated as t- or g-matrix  
at starting energies $E_\alpha = E+\epsilon_\alpha$ in the 
NA CM frame (subscript A+N) \cite{Are95}. In the absence of medium 
modifications or at high energies, the g-matrix is approximated 
by the t-matrix. Since medium modifications have been shown to 
be important only for energies $E<700$\,MeV we will
make our relativistic corrections as if the effective 
interaction $\cal T$ is a t-matrix. 

The optical potential Eq.\,(\ref{eqn_III.1}) requires the 
NN t-matrix in the projectile-nucleus CM frame. 
Generally, the t-matrices are calculated  in the NN CM system  
with a Lippmann--Schwinger equation and a two nucleon  potential;
the practical problem is how to translate the t-matrix into the proper
reference frame. This problem is not new and has been discussed 
elsewhere \cite{Aar68,Ern80,Gie82}. They 
use simply a Lorentz boost for the kinematic variables between 
the reference frames.

In impulse approximation, the transformation of the t-matrix from the 
NN to the NA  CM frame can be done sequentially.
First, all four momenta in the NA CM system need to be transformed to 
the NN CM system.
Second, as Lorentz invariance of the flux is assumed, the M\o ller 
flux factor transforms  the NN t-matrices. 
Third, a spin rotation between the reference systems
requires a Wigner rotation \cite{Tjo91}.
Guided from Fig.\,\ref{Figure_A1}, the NN
interaction in Eq.\,(\ref{eqn_III.1}) involves incoming and 
outgoing four-momenta
$k=(\omega,{\bf k})$ of the projectile and struck-nucleon 
$p=(\epsilon,{\bf p})$ respectively.
The effective  NN interaction is translated by 
\begin{equation} \label{eqn_III.2}
\left \langle {\bf  k}^\prime {\bf p}^\prime |{\cal T}(E_\alpha)| 
{\bf k} {\bf  p }\right \rangle_{A+N} = 
\eta(k^\prime\,  p^\prime | k\, p ) \,
\left \langle \kappa^\prime,-\kappa^\prime
| t_{Q }(E_\alpha)| \kappa , - \kappa\right \rangle_{NN} \;
\delta ( \vec Q^\prime - \vec Q ) 
\end{equation}
with the M\o ller factor
\begin{equation}
\label{eqn_III.3}
\eta(k^\prime\,  p^\prime | k\, p)=
\sqrt{ {\omega(\kappa') \epsilon(\kappa') \omega(\kappa)\epsilon(\kappa)
\over \omega(k') \epsilon(p') \omega(k)\epsilon(p)}}.
\end{equation}

The relativistic kinematics of Aaron, Amado and Young \cite{Aar68} and 
Giebink's \cite{Gie82} are used  in the calculation  of the 
full-folding approach and  details are found elsewhere \cite{Are01}.

\begin{figure}[H]
\centering
\epsfig{file=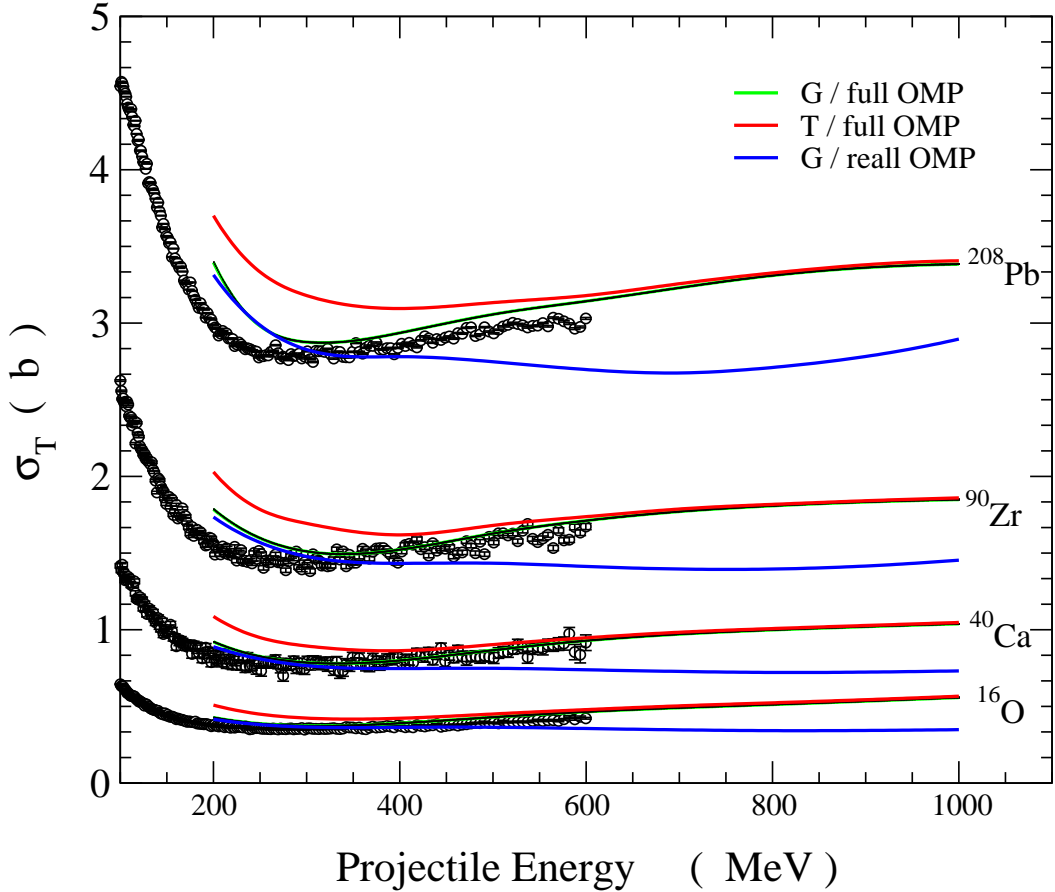,scale=0.90}
\caption{Total cross sections for A(n,n)A based upon the 
Argonne AV18 reference and 
full NN  optical model g-matrix (green) and t-matrix (red); and 
the Argonne (AV18) reference and real part NN optical model 
g-matrix (blue)}
\label{Figure_N10}
\end{figure}

\begin{figure}[H]
\centering
\epsfig{file=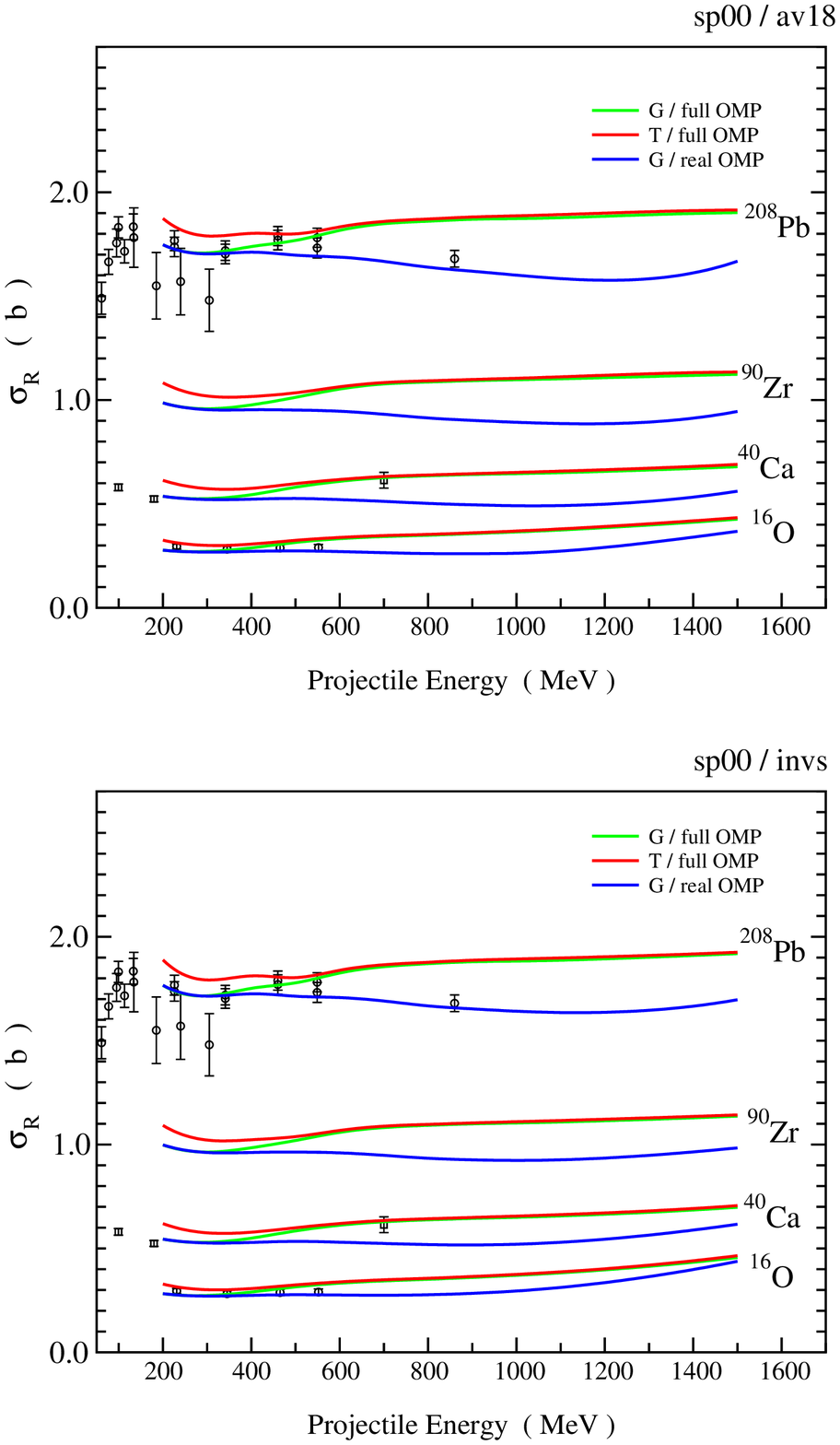,scale=0.75}
\caption{Reaction cross sections for A(n,n)A based upon the 
Argonne AV18  and inversion INVS reference and 
full NN  optical model g-matrix (green) and t-matrix (red); and 
the Argonne (AV18)  and inversion INVS reference and real 
part NN optical model g-matrix (blue).}
\label{Figure_N9}
\end{figure}

An assessement of the sensitivity of the full-folding optical potential
to the inclusion of relativistic kinematics over a wide energy range 
can be made by studying reaction and total  cross sections 
for neutron elastic scattering.
In Figs. \ref{Figure_N9} and \ref{Figure_N10} 
we show the measured \cite{Fin93} and calculated reaction 
and total cross sections 
for neutron elastic scattering from $^{16}$O, $^{40}$Ca, $^{90}$Zr
and $^{208}$Pb at projectile energies from 100 MeV up to 1.5 GeV. 
These cross sections are obtained by solving the scattering problem 
using full-folding optical potentials calculated as in \cite{Are95},
with the kinematics discussed in the most recent work \cite{Are01}. 
 
As a very important comparison,  we have included results using 
{\em medium independent} t-matrix
and  {\em medium dependent} g-matrix  effective interactions. 
These calculations confirm the need and superiority 
of g-matrix effective interactions over 
the entire energy range.
Furthermore, the description of reaction and total cross sections  by the
relativistic full-folding optical model is quite remarkable.
Some difficulties are noticeable for  data 
$<300$\,MeV, where the calculated values are above the data. 
This does not manifest a principle
failure of the full-folding model
but would require attention. To this end we feel obliged to point to 
some remarkable results of  microscopic optical 
model calculations \cite{Deb01}.

A fundamental understanding of the structure of 
nuclei from a QCD point of view
and the interest to understand the behavior of nucleons within nuclei
is an issue. Total and reaction cross sections for nucleon  
elastic scattering from low to high 
energies have practical use in the operation of spallation and other nuclear 
facilities \cite{Kal96,Dub97,Mas98,Rip98}.
Thus, we show our ability to calculate  total and reaction  
cross sections for elastic nucleon (shown for neutron) scattering at 
energies $<1.5$\,GeV. In Figs. \ref{Figure_N10} 
and \ref{Figure_N9} are displayed g-matrix 
(green), and t-matrix (red) results based upon
the Argonne (AV18) and our inversion (INVS) reference and 
complex separable NN optical potentials, including data from \cite{Fin93}.
The AV18 and INVS reference potentials and real part separable NN optical potential 
was used for a g-matrix calculation (blue). 
The calculations serve at least two purposes. The first
is an assessement of the  importance of 
NN inelasticities (imaginary part of NN OMP)
in comparison with the NA inelasticity (contains both, the NN optical
model and NA pole term). The calculations 
give a quantitative impression about this effect and its energy dependence.
The opening of NN inelasticities becomes visible at $T_{Lab}\sim 400$\,MeV
and contributes $\sim 20$\,\% of the total and reaction
cross sections at higher energies 1-1.5\,GeV for all isotopes.
The other aim of the study is to show that the NN full optical 
model  g-matrix calculations are superior to 
t-matrix calculations for $300< T_{Lab} < 700$\,MeV. 
It is not unexpected but important to know that the choice of reference potential
has some  effect but the differences between AV18 and INVS
are visible in Fig. \ref{Figure_N9}. This confirms
the importance and need of on-shell equivalent potentials and careful 
inclusion of off-shell g- and t-matrices at any stage of 
a folding calculation \cite{Amo00}.
The calculated results also show a nearly constant cross section
as function of  energy above 1 GeV. We are not aware of data above
1 GeV to confirm that part of our calculations.  
The A(n,n)A differential cross section and spin observables
\cite{Are01} give  results which imply the same conclusions as drawn ffrom the
total and reaction cross sections. Comprehensive studies have been performed
for the nuclei $^{16}$O, $^{40}Ca$, $^{90}$Zr and $^{208}$Pb.
Here again we observe hardly  differences between the g- and t-matrix
approaches at energies above 700 MeV but the differences become
pronounced at  lower energies.

A few remarks are noteworthy regarding the full-folding applications
made here as compared with calculations made in the 
context of the nonrelativistic impulse approximation t$\rho$ 
based on the Kerman, McManus and Thaler formalism \cite{KMT59}.
The calculation of optical potentials within the full-folding approach 
presented here
are made without assumptions neither about the local structure of the NN
effective interaction nor the final structure of the NA coupling.
In fact, these potentials are treated as non-local potential 
operators obtained 
from a detailed account of the NN effective interaction off-shell. 
In contrast, the momentum space  optical potentials in 
the nonrelativistic impulse 
approximation are often  assumed as being  local and in  
calculations only on-shell g- and t-matrix  elements of the 
NN effective interaction enter. 
Important differences have been observed between these two 
approaches when 
applied to intermediate energy (200-400 MeV) NA scattering 
\cite{Ray92}.
Both full-folding and t$\rho$ approximation
calculations were made using the free t-matrix
with Giebink relativistic kinematics.
Differences between these two approaches become noticeable
for momentum transfers $q > 1$\,fm$^{-1}$ and  the importance of 
off-shell effects are obvious.
The extent of this sensitivity is comparable to 
contributions from short-range correlations. 
The most significant difference between  results of the past and the
ones presented here are coming from the NN optical 
model potential used.
Medium effects due to Pauli blocking and mean 
field effects are  small for energies above 700 MeV.
A definitive signal of NN dibaryonic formation in NA scattering could not
be identified as yet. 

\end{document}